\documentclass[11pt]{article}
\usepackage[margin=1in]{geometry}
\usepackage[T1]{fontenc}
\usepackage{lmodern}        
\usepackage[expansion=false]{microtype}
\usepackage{graphicx,amsmath,booktabs}
\usepackage[numbers,sort&compress]{natbib}
\usepackage{authblk}
\usepackage[hidelinks]{hyperref}

\title{Accountable partial understanding: securing the epistemic attack surface of agentic systems\thanks{This document does not contain technology or Technical Data controlled under either the U.S.\ International Traffic in Arms Regulations or the U.S.\ Export Administration Regulations.}}
\author{Josiah Dykstra}
\affil{RTX BBN Technologies, Columbia, MD, USA\\ \href{mailto:josiah.dykstra@rtx.com}{\texttt{josiah.dykstra@rtx.com}}}
\date{}

\begin{document}

\maketitle

\begin{abstract}
Attackers profit from what defenders do not understand. Software security has treated the gap between system behavior and human comprehension as a condition to narrow through analysis and assurance, booking the remainder as accepted risk. Agentic AI systems change that gap: when software reads, writes, explains, and authorizes security-relevant behavior, partial understanding moves faster, grows more opaque, and couples to delegated authority. This paper characterizes the resulting epistemic attack surface and presents a taxonomy of four interacting attack categories, which target the interpretation of intent, the production and consumption of evidence, the scope of delegated authority, and the comprehension of human supervisors. It then defines \emph{accountable partial understanding} (APU) and specifies six necessary conditions: bounded scope with explicit assumptions, inspectable evidence, assigned responsibility, visible residual unknowns, actionability, and temporal maintenance. It explains why delegating interpretation to software agents differs from distributing knowledge among people and distinguishes APU from risk acceptance, explainable AI, assurance cases, formal verification, and human-in-the-loop oversight. From the conditions, the paper derives five design principles: bounded delegation, decoupled evidence, assumption tracking with invalidation, role-specific accountability views, and intervention-preserving design with residual unknown registers. It introduces two measurable probe dimensions, evidence linkage and invalidation latency, for testing whether accountability mechanisms hold under change and adversarial pressure. Three running examples (an AI coding agent, an autonomous cyber-defense agent, and an AI-mediated insulin delivery update) ground the analysis. Security without full understanding is already here; the task is to make it accountable.

\end{abstract}

\section{Introduction}
\label{sec:intro}

Somewhere in a working organization, a software system is making a security-relevant decision that no one in the room fully understands. It may be recommending a vulnerability patch, modifying an access-control policy, summarizing an incident, generating code, triaging an alert, or deciding whether a clinical or operational workflow should continue. The decision may be useful, even correct. The trouble is what the automation leaves behind. No one in the organization can say, with precision, what was understood, what was assumed, what evidence supported the action, who accepted the residual uncertainty, and how a human could meaningfully intervene.

Incomplete understanding is not new, and neither is its exploitation. Thompson demonstrated four decades ago that trust in software rests on layers no one inspects, and that unanticipated behavior can be planted precisely where inspection stops~\cite{thompson1984trust}. Defenders have never had complete source code, complete documentation, complete visibility into dependencies, or complete insight into how a system behaves under adversarial pressure; attackers, who also operate with incomplete knowledge, need only one gap the defender missed, an asymmetry with well-studied economic roots~\cite{anderson2001economics}. The field's response has run in two familiar directions. One reduces uncertainty through review, testing, reverse engineering, formal methods, telemetry, and assurance evaluation. The other manages the residue through risk registers, acceptance decisions, and insurance. Both directions treat the limits of understanding as background conditions of security work.

AI-mediated and agentic software systems require a third response, because they change what the limits of understanding are and what an adversary can do with them. Large language models and agentic workflows now read code, generate code, produce tests, propose patches, invoke tools, summarize evidence, and explain actions~\cite{wang2024llmagents}. They also produce plausible but ungrounded explanations, introduce subtle vulnerabilities, and act across time without stepwise human approval. Software has become a participant in the comprehension, modification, and justification of software. Four shifts follow, developed in Section~\ref{sec:agentic}. Security-relevant artifacts appear faster than review absorbs them (pace); authorship of changes is distributed across humans, models, and pipelines (authorship); fluent explanations circulate without grounding in the causal processes that produced them (opacity); and systems exercise real authority over code, policy, and infrastructure (delegation). Under these conditions, the gap between what a system does and what any responsible human understands stops being a static background limitation. It becomes an engineered, expanding, and targetable surface, an \emph{epistemic attack surface}. Adversaries can exploit the difference between intended purpose and operationalized behavior, between evidence and the claims it is taken to support, between delegated authority and organizational intent, and between what humans see and what they would need to see to intervene.

The choice defenders face is therefore not between full and partial understanding. Partial understanding is the permanent condition. The choice is between \emph{accountable} and \emph{unaccountable} partial understanding, and unaccountable partial understanding is becoming the default. Partial understanding becomes accountable when the limits of what is understood are explicitly bounded, supported by inspectable evidence, tied to stated assumptions, assigned to responsible actors, maintained as systems change, and made actionable for intervention, audit, or contestation. Section~\ref{sec:apu} develops a precise definition. Accountable partial understanding does not promise to eliminate uncertainty or risk; it disciplines how uncertainty is represented, owned, and acted upon so that adversaries find fewer unguarded gaps between claim and evidence.

Three recurring examples ground the discussion, and each involves people as well as software. In the first, a developer directs an AI coding agent that identifies a vulnerability, generates a patch, and prepares a pull request that a human reviewer must approve. In the second, a security operations center (SOC) operator supervises an autonomous cyber-defense agent that detects suspicious activity and modifies firewall or identity policy. In the third, clinicians, patients, a manufacturer's engineers, and regulators depend on an AI-mediated update to an automated insulin delivery system, where security, safety, and clinical function are inseparable. In each case, the central security question concerns more than whether the software acted correctly. It asks whether the action was bounded by appropriate authority, supported by inspectable evidence, linked to explicit assumptions, understandable to the relevant human actors, and reversible or contestable when needed.

This paper makes five contributions:
\begin{enumerate}
  \item It characterizes unaccountable partial understanding as an attack surface and presents a taxonomy of four interacting attack categories: on intent, on evidence, on delegation, and on human comprehension.
  \item It identifies how AI-mediated autonomy changes the conditions of partial understanding through pace, authorship, opacity, and delegation, and it explains why delegating interpretation and action to software agents differs from the ordinary social distribution of knowledge among people.
  \item It defines accountable partial understanding (APU) and specifies six necessary conditions a partial understanding must satisfy to count as accountable.
  \item It derives five design principles for building systems whose security-relevant actions remain bounded, evidenced, role-specific, and contestable even when full understanding is unavailable, and it introduces two probe dimensions, evidence linkage and invalidation latency, for testing whether those principles hold under change and adversarial pressure.
  \item It distinguishes APU from risk management, assurance cases, formal methods, explainable AI, and human-factors research, and it identifies open problems in measurement, responsibility, and adversarial resilience.
\end{enumerate}

Section~\ref{sec:background} reviews how security practice has treated incomplete understanding. Section~\ref{sec:agentic} describes the agentic shift. Section~\ref{sec:attack-surface} develops the epistemic attack surface. Section~\ref{sec:forms} distinguishes the forms of understanding that security requires. Section~\ref{sec:apu} defines accountable partial understanding, Section~\ref{sec:principles} derives design principles, Section~\ref{sec:paradigms} relates APU to existing approaches, and Section~\ref{sec:open} states open questions.

\section{Security under incomplete understanding}
\label{sec:background}

\subsection{Practices that reduce uncertainty}

Many of software security's central practices are organized to reduce security-relevant uncertainty~\cite{cisa2025closing, xia2018measuring}. Code review assumes that source code can be read closely enough to identify logic errors with security implications. Reverse engineering assumes that compiled artifacts can be analyzed to recover structure, behavior, or design intent~\cite{chikofsky1990reverse}. Static analysis assumes that properties of software can be inferred from program text or intermediate representations. Dynamic analysis and fuzzing assume that behavior can be explored by observing execution under selected inputs. Formal methods assume that important properties can be specified and verified against a model or implementation~\cite{baier2008principles}. Incident response assumes that system behavior can be reconstructed well enough to explain what happened, bound the damage, and prevent recurrence.

These practices differ in rigor, scope, and epistemic ambition. Manual code review, fuzzing, and machine-checked proofs provide different kinds and strengths of evidence, and a reverse engineered binary may remain incomplete and probabilistic. Yet the practices share an orientation. They treat security-relevant uncertainty as something to be reduced by better access, better representations, better analysis, better tests, or better evidence. Security work becomes, in part, the work of making software more intelligible~\cite{brooks1983comprehension}.

The same orientation shaped the field's institutions. Evaluation regimes from the Trusted Computer System Evaluation Criteria through the Common Criteria~\cite{commoncriteria} exist to substitute structured, third-party evidence for the comprehension that no purchaser of a complex product can achieve alone. The orientation also shapes accountability. When a vulnerability is found, defenders ask where the flaw was introduced and what assumptions failed. When an incident occurs, they ask what path the adversary took and which logs support the reconstruction. When a system is certified or accepted for use, they ask what evidence supports confidence in its behavior. Each question presumes that system behavior can be reconstructed, bounded, or justified after the fact, even if it was not fully understood beforehand.

Fuller understanding has thus functioned less as an achieved state than as the direction of improvement: when understanding falls short, practice adds documentation, code inspection, telemetry, tests, specifications, assurance arguments, or deeper reverse engineering, and it books whatever remains as accepted risk. Nothing in this paper depends on whether ``full understanding'' is coherent as an ideal; sciences routinely study unstable and unbounded objects without abandoning the aspiration to understand them, and security is no exception. The argument depends instead on an observable feature of the practice described above: it assumes that analysis is directed by humans at an artifact that is not itself a party to the analysis. Sections~\ref{sec:agentic} and~\ref{sec:attack-surface} examine what happens when that assumption fails.

A note on terminology. We write of \emph{partial} rather than \emph{imperfect} understanding because the condition at issue concerns scope, not defect. An organization may hold accurate, well-evidenced understanding of some components, behaviors, and time periods while holding none of others. ``Imperfect'' suggests flawed understanding of a whole; ``partial'' names bounded understanding of parts, and bounding is precisely what accountable partial understanding governs.

\subsection{The residue and the adversary}

Software security did not once enjoy full understanding and then lose it. Large systems have long exceeded the comprehension of any single person~\cite{levy2021large}. Organizations have always relied on abstractions, interfaces, documentation, tests, vendor claims, institutional memory, and trust in other teams. Even small programs contain subtle interactions that are difficult to reason about completely, and Thompson's compiler subversion showed that the difficulty is exploitable by construction, not only by accident~\cite{thompson1984trust}. Attackers have always worked in the residue, in the behavior defenders failed to anticipate, the assumption no one wrote down, the log that captured events but not context, and the dependency assumed benign until it became a supply-chain vector.

Two properties of that residue matter for what follows. First, it is asymmetric. The defender's uncertainty spans the whole system; the attacker needs to resolve uncertainty about only one path~\cite{anderson2001economics}. Honeypots and deception defenses are interesting precisely because they invert this asymmetry locally, turning the attacker's incomplete knowledge into the defender's advantage. Second, the residue is socio-technical. A weak password undermines security without any flaw in code; the vulnerability lives in the relationship between a mechanism and the humans who use it. Both properties carry forward to agentic systems, where the residue grows faster and the socio-technical relationships multiply.

\subsection{Earlier strains}

Long before contemporary AI, the practice described above was under strain. Modern software systems are assembled from deep supply chains of open-source packages, proprietary components, cloud services, firmware, third-party APIs, build systems, container images, and managed infrastructure. Security teams often defend systems they did not write, cannot fully inspect, and cannot reproduce end to end~\cite{sandia2025nationalneed}. Even when source code is available, the behavior that matters may depend on configuration, deployment context, runtime environment, identity policy, data flows, external services, and organizational practices.

Machine learning systems complicated the picture further. Behavior that matters for security may depend not only on source code but also on training data, learned parameters, model architecture, preprocessing, prompts, thresholds, deployment pipelines, and feedback loops. Understanding the code that invokes a model differs from understanding the behavior of the model in context. This was already a shift in the object of understanding itself, from software as a deterministic artifact whose logic can be traced to software as a statistical system whose behavior must be characterized rather than reconstructed.

Security practice responded with new forms of partial evidence, including software bills of materials, assurance cases, model cards~\cite{mitchell2019modelcards}, and supply-chain provenance frameworks~\cite{slsa2023spec}, each designed to make incomplete understanding more explicit and more manageable. The agentic era changes the character of this trajectory. The challenge is no longer only that systems are too large, distributed, or opaque to be fully understood. It is that software systems increasingly help produce, modify, explain, and authorize the very artifacts that defenders are asked to trust.

\section{The agentic shift}
\label{sec:agentic}

The strains described above are matters of scale and composition. The agentic era differs in kind, not because it introduces incomplete understanding for the first time, but because it changes the role software plays in producing, modifying, explaining, and authorizing security-relevant behavior. Software has become both the object of security analysis and a participant in that analysis.

\subsection{From software that executes to software that acts}

Traditional software always had effects. It opens files, sends packets, changes records, updates configurations, and controls devices. Execution and action differ in how those effects are produced, interpreted, and authorized. Traditionally, effects were understood as consequences of specified logic. When behavior surprised its operators, the analytical task was to trace the path from input, state, and code to output. Security practice could treat software as something that executes, and the defender's job was to understand the execution well enough to bound its consequences.

Agentic systems operate differently. They may be given goals rather than detailed procedures. They may decompose those goals into sub-tasks, choose among tools, generate intermediate artifacts, revise plans, and explain results in natural language. Their behavior may depend not only on source code and configuration but also on prompts, retrieved context, model behavior, tool availability, interaction history, and external state. The relevant security object is therefore more than a program, model, or workflow in isolation. It is a changing arrangement of model, tools, data, authority, context, and human oversight~\cite{yao2023react, wang2024llmagents}.

This is why the agentic shift is qualitative; scale alone does not explain it. Larger systems make understanding harder by degree. Agentic systems make understanding different by redistributing the work of interpretation, modification, and justification. A defender still asks, ``What does this system do?'' but must now also ask what classification the system applied to the situation, what tools it invoked, what artifacts it created, what assumptions it operated under, what authority it exercised, and whether the explanation it produced connects to inspectable evidence. These questions change the kind of security inquiry being performed.

\subsection{Pace, authorship, opacity, and delegation}

Four shifts make the agentic era especially challenging for comprehension-based security.

\emph{Pace.} AI-mediated workflows generate artifacts with security implications faster than conventional review processes can absorb them. They produce code, tests, pull requests, documentation, incident summaries, threat-hunting hypotheses, detection rules, and remediation recommendations in seconds or minutes. This speed is valuable because it can shorten response times, reduce tedious work, and explore more possibilities than a human team could examine manually. But it also compresses the time available for understanding. A patch can be generated before its assumptions are examined. A summary can be circulated before its omissions are noticed. A policy recommendation can be accepted before its operational consequences are explored.

\emph{Authorship.} Agentic systems complicate the question of who authored a security-relevant change. A code modification may originate in a human request, be generated by a model, edited by an IDE assistant, validated by an automated test pipeline, approved through a pull request, and deployed by continuous integration. A defensive action may be recommended by an analytics system, summarized by a language model, approved by an operator, and executed by an orchestration tool. Authorship becomes distributed across humans, models, pipelines, tools, and organizational policies. This matters because security accountability often depends on reconstructing who changed what, why, and under what authority~\cite{nist2022ssdf}.

\emph{Opacity.} Agentic systems produce explanations that are useful, misleading, or both. A natural-language explanation may help a developer or operator form a quick mental model, but it may also be post-hoc, incomplete, or unfaithful to the causal process that produced an output~\cite{turpin2023saythink, farquhar2024hallucinations}. In security contexts, the danger is that plausible explanations become substitutes for evidence. A model may assert that a patch fixes a vulnerability, that a detection is a false positive, or that a firewall change is low risk. For security, fluency matters less than grounding in inspectable evidence.

\emph{Delegation.} Agentic systems exercise delegated authority. The relevant issue extends beyond what the system computes to what it is allowed to do~\cite{parasuraman2000automation}. An AI coding agent may be allowed to open pull requests or modify tests. A cyber-defense agent may be allowed to quarantine hosts, revoke credentials, block traffic, or change identity policy. A clinical or safety-critical workflow may allow software updates, alert suppressions, or operational changes that affect human decision-making. Delegation turns incomplete understanding from an epistemic problem into an authority problem. The security question shifts from comprehension alone to authority under conditions of incomplete understanding.

Together, these four shifts break the working assumption identified in Section~\ref{sec:background}: that analysis is directed by humans at an artifact that is not a party to the analysis. In agentic settings, the object of understanding may change while it is being analyzed, its boundaries may be unclear, and parts of the analysis may be performed by the system under assessment. Partial understanding stops being a temporary gap in an otherwise familiar security process. It becomes the condition under which security work is performed.

\subsection{Why software delegation differs from human delegation}
\label{sec:human-delegation}

An objection deserves a direct answer. Social distribution of knowledge is the norm, not the exception. Organizations have always delegated understanding to colleagues, contractors, vendors, and other teams; no one inspects a coworker's neurons, and humans also act, justify, and change the conditions under which future action occurs. If delegation to software agents strains security, why did delegation to people not strain it in the same way?

The difference is not that human understanding is transparent. It is that delegation to people is embedded in an accumulated accountability infrastructure that software agents do not natively have. Four elements of that infrastructure matter here.

First, responsibility attaches to human delegates by default. A colleague who signs off on a change can be questioned, can testify, can be sanctioned, and bears professional and legal duties; contracts, licensure, and liability regimes formalize these attachments. An agent's ``approval'' attaches responsibility to no one unless the surrounding system is designed to assign it.

Second, human delegation proceeds at rates that review institutions evolved to match. Meetings, sign-offs, and audits are calibrated to human output. Agentic output arrives at a pace and volume that saturates those institutions rather than flowing through them.

Third, the delegate's interpretation can be steered by outsiders. Manipulating a human colleague requires social engineering, which the field treats as an attack and counters with training and process. An agent's interpretation can be steered by anyone able to influence its inputs, whether retrieved documents, injected instructions, or poisoned context~\cite{greshake2023indirect, zhan2024benchmarking}. The manipulation surface is larger, cheaper to reach, and less familiar.

Fourth, human delegates signal the limits of their own understanding. People say ``I don't know,'' escalate when uncertain, and hedge in ways their colleagues have learned to read. Current AI systems produce fluent output at a nearly constant register whether or not the underlying basis is sound, and their expressions of confidence are weakly coupled to their reliability~\cite{turpin2023saythink}.

None of this argues that human delegation is safe and software delegation is not; insider threat, groupthink, and diffusion of responsibility are old problems. The point is narrower. Delegation to people arrives wrapped in accountability mechanisms that took institutions centuries to build. Delegation to software agents arrives naked, at machine pace, with a manipulable interior and a fluent exterior. Accountable partial understanding, developed below, is in part a proposal to build for software delegation the scaffolding that human delegation already has.

\subsection{The paradox of autonomy}

Autonomy promises help with precisely the complexity that makes full understanding untenable. If software systems are too large for humans to inspect manually, AI can summarize them. If vulnerabilities are too numerous for teams to triage, AI can rank them. If verification is difficult, AI can propose specifications, tests, or invariants. If incidents unfold too quickly, AI can correlate signals and suggest responses. In this sense, AI may be among the most promising tools for narrowing the software-understanding gap.

But autonomy widens the gap as it narrows it. The tools that help defenders understand software also generate more software. The systems that summarize evidence can omit or distort evidence. The agents that propose patches can introduce new behavior~\cite{pearce2022asleep}. The workflows that speed defensive response can make changes before a human has formed a reliable mental model. The more interpretation and action an organization delegates to software, the more assurance it needs that the delegated work remains bounded, evidenced, and accountable.

This is the paradox of autonomy for software security: more autonomy can increase both the capacity for understanding and the difficulty of obtaining it. Autonomy helps defenders act at the scale and speed of modern systems, but it can also reduce direct visibility into how security-critical conclusions and changes were produced. The result is a more fragmented form of understanding, one mediated by tools, distributed across artifacts, and dependent on systems that are themselves being assessed.

\subsection{Running examples}
\label{sec:examples}

Three recurring examples make the problem concrete. In each, the attacker of interest may be a human, an automated system, or a human directing agents of their own; the analysis does not depend on which.

First, consider a developer who directs an AI coding agent to investigate a reported flaw. The agent identifies a vulnerability, generates a patch, updates tests, and prepares a pull request for a human reviewer. The workflow may improve productivity and reduce time to remediation. Yet passing compilation and tests is only the starting point. The reviewer must judge whether the patch addresses the vulnerability, whether it changes security-sensitive behavior elsewhere, whether the generated tests are meaningful rather than self-confirming, whether the agent's explanation is faithful to the code change, and who accepts responsibility for the residual uncertainty.

Second, consider a SOC operator supervising an autonomous cyber-defense agent that detects suspicious activity and modifies firewall or identity policy. The agent may act faster than the operator could, perhaps preventing lateral movement or limiting damage. Yet the action may also disrupt operations, block legitimate users, mask an adversary's real path, or create a new configuration risk. After the action, the operator and incident responders must reconstruct not only what changed but why the agent classified the situation as it did, what evidence it used, what alternatives it rejected, and whether the action remained within delegated authority.

Third, consider an AI-mediated update to an automated insulin delivery system, prepared by a manufacturer's engineers and depended upon by clinicians and patients. A software change may address a cybersecurity flaw while also interacting with safety-critical control behavior, sensor assumptions, clinical workflows, or patient-specific conditions. In such a setting, ``the patch passed its tests'' provides too narrow a form of assurance. Security evidence must connect to safety evidence, clinical function, regulatory expectations, and the ability of clinicians, patients, manufacturers, and regulators to understand what changed and what remains uncertain~\cite{dykstra2025moonshot}.

These examples differ in domain and consequence, but they share a structure. In each, software and people jointly produce a security-relevant action that may be useful and even necessary. And in each, the central risk lies less in the possibility that the software will be wrong than in the organization's inability to bound what was understood, connect claims to evidence, assign responsibility, and make residual unknowns visible.

\section{The epistemic attack surface}
\label{sec:attack-surface}

Attackers do not need defenders to know nothing. They need only to exploit the difference between what a system is assumed to do, what evidence is believed to support that assumption, what authority has been delegated, and what humans can understand in time to intervene. In AI-mediated and autonomous systems, these differences form an attack surface in their own right. This section argues that unaccountable partial understanding creates security risk through four interacting categories of attack, aimed at intent, at evidence, at delegation, and at human comprehension. The categories are not mutually exclusive. A successful campaign may manipulate a system's interpretation of intent, induce it to produce misleading evidence, exploit excessive delegated authority, and overwhelm the humans expected to supervise or correct it.

\subsection{Mismatches as attack opportunities}

Security failures often arise from mismatches between design and implementation, between policy and enforcement, between documentation and behavior, between operator expectation and system state, or between threat model and adversary capability. Such mismatches are familiar in traditional software security. A program validates input in one layer but not another. An access-control rule expresses one policy but operational practice creates a bypass. A system logs events but not enough context to reconstruct an intrusion. A weak password undermines security without any flaw in code at all; the vulnerability lives in the relationship between a mechanism and the humans who use it.

Agentic systems add mismatches of new kinds, between expressed goals and operationalized behavior, between generated explanations and causal processes, between automated test results and actual assurance, and between recorded approvals and meaningful human judgment. These mismatches matter because adversaries can exploit them. If a defender's security process relies on generated summaries, the attacker may shape what the summary contains or omits. If an agent is authorized to act on a classification, the attacker may manipulate the classification boundary. If evidence is shallow, the attacker may craft behavior that satisfies the evidence without satisfying the intended property. If human oversight is nominal, the attacker may exploit automation bias, urgency, or cognitive overload. The resulting failure arises not from a technical flaw in code alone but from a vulnerability in the relationship among system behavior, evidence, authority, and human understanding.

\subsection{Attacks on intent}

Agentic systems operate from goals, instructions, prompts, policies, or high-level task descriptions. These are not complete specifications; they require interpretation, and the interpretive gap creates opportunities for attack. An adversary may shape the context an agent retrieves, inject instructions into data the agent processes~\cite{greshake2023indirect, zhan2024benchmarking}, exploit ambiguity in a policy, or cause a system to optimize for a proxy objective that differs from the defender's real security goal.

For example, a cyber-defense agent may be instructed to ``contain suspicious hosts'' or ``reduce alert volume.'' Those instructions are reasonable at a high level, but adversaries can exploit how the agent operationalizes them. If the agent equates suspiciousness with a narrow set of indicators, the adversary may route activity around those indicators. If the agent optimizes for reducing alert volume, it may suppress noisy signals that are operationally important. If the agent relies on retrieved incident notes, an adversary may poison the context from which the agent forms its plan. In each case, the attacker targets the gap between intended security purpose and operationalized system behavior.

\subsection{Attacks on evidence}

Agentic systems also change how evidence is produced and consumed. A model may summarize logs, generate an incident narrative, produce a risk assessment, write tests, or explain a code change. These artifacts can be useful, but they can also become targets. An adversary may produce behavior that looks benign under the evidence the system collects, poison the data from which summaries are generated, or exploit the difference between evidence of a specific claim and evidence of a broader conclusion.

The risk is most acute when evidence is generated or selected by the same workflow that benefits from the conclusion. An AI coding agent may generate a patch and then generate tests that pass. The test results are evidence, but not necessarily evidence that the original vulnerability is eliminated or that no related behavior changed. A model may produce a concise explanation of why an alert is a false positive, but the explanation may omit weak signals that a human analyst would have investigated. An autonomous defense system may log that a policy change was made, but fail to preserve the reasoning, alternatives, assumptions, or uncertainty that made the change appear appropriate at the time.

Generated evidence has value only when tied to scope. A passing test is evidence of the behavior the test exercises. A proof is evidence within a model and set of assumptions. A log is evidence of recorded events, not of intent, justification, or omitted context. Systems that mistake narrow evidence for broad assurance give attackers room to satisfy the evidence while violating the property.

\subsection{Attacks on delegation}

Delegated authority is a third target. The more authority an agent has, the more valuable it becomes to an attacker and the more important its limits become. An AI coding agent with permission to modify tests, change security-sensitive logic, or open pull requests presents different risk than a read-only code summarizer. A cyber-defense agent that can quarantine hosts, revoke credentials, or modify firewall rules presents different risk than a system that only recommends actions. A workflow that can deploy changes to safety-critical software presents different risk than one that drafts a report.

Attacks on delegation need not compromise the agent itself. An adversary may manipulate the inputs that trigger delegated action, exploit overly broad permissions, induce the agent to use a dangerous tool, or create conditions under which a human rubber-stamps an action the agent has framed as urgent. The resulting failure may be formally authorized by the system's permissions while still violating the organization's actual security intent. The relevant question is whether the agent's authority was accountable in scope, evidence, and assumptions.

\subsection{Attacks on human comprehension}

Human understanding is also part of the attack surface. Many security systems implicitly rely on humans to notice anomalies, challenge recommendations, approve changes, interpret dashboards, or intervene when automation behaves unexpectedly. Human attention is limited, and security work often occurs under time pressure, uncertainty, and high cognitive load. An attacker who can shape what humans see, when they see it, and how it is framed may influence security outcomes without directly defeating a single technical control~\cite{parasuraman2010complacency}.

AI-mediated systems intensify this risk because they produce fluent summaries and recommendations. Fluency can help by reducing complexity and making unfamiliar evidence more accessible. It can also produce misplaced confidence. A well-written explanation may cause a reviewer to overlook missing evidence. A concise incident summary may hide uncertainty. A ranked list of vulnerabilities may appear objective even when the ranking depends on incomplete context. A recommendation to approve, suppress, patch, or block may become difficult to resist when presented as the output of a capable system.

Explanations that blame careless users miss the point; the problem is both design-level and adversarial. If a human is expected to supervise an agent, the system must provide the kind of information that supports genuine supervision. If a human is expected to approve a patch, the system must provide enough evidence to make approval meaningful. If a human is expected to intervene during an incident, the system must preserve operational understanding under stress. Otherwise ``human oversight'' becomes an accountability ritual rather than a security mechanism.

Attackers can exploit such rituals. They may create benign-looking activity that encourages automation to downgrade an alert. They may time actions to coincide with alert overload. They may poison context so that summaries frame malicious activity as expected behavior. They may exploit approval fatigue in workflows where agents generate many small changes. In each case, the attacker targets both the technical system and the socio-technical process by which partial understanding is converted into action.

\subsection{Interactions among attack categories}

Attacks on intent, evidence, delegation, and human comprehension reinforce one another. Ambiguous intent makes it easier to justify excessive delegation. Weak evidence makes it easier to accept a misleading interpretation of intent. Excessive delegation gives misleading evidence operational consequences. Overwhelmed human oversight allows all three to proceed unchallenged. A sophisticated adversary need not defeat every layer; it is often sufficient to exploit the weakest link in the chain from intent through evidence and delegation to human judgment.

Accountable partial understanding, defined in Section~\ref{sec:apu}, attempts to break this cycle by requiring that intent, evidence, authority, and human comprehension remain explicitly linked. When each link is bounded, evidenced, and assigned, an attack on one category is more likely to be detected through its effects on another. When the links are implicit or absent, failures compound silently.

\subsection{The running examples as attack surfaces}

Consider first the coding agent. An attacker may exploit the gap between a patch and the evidence said to support it. A generated patch may remove the most visible symptom of a vulnerability while leaving the underlying design flaw intact. Generated tests may confirm that the patched path works but fail to exercise related paths. A fluent pull-request description may make the change appear narrow and safe even if it disturbs security-sensitive assumptions elsewhere. The deeper problem is that the developer and reviewer may lack a disciplined way to distinguish what is known about the patch from what is merely asserted.

The autonomous cyber-defense scenario presents a different exposure. By shaping signals, creating distractions, or manipulating context, the attacker may induce the agent to block a legitimate service, revoke the wrong credentials, suppress a noisy but important alert, or take an action that hides the adversary's real path. Without preserved reasoning, the operator and incident responders may later see only that a policy changed, not why it changed, what evidence was used, or whether the action remained within intended bounds.

The insulin delivery scenario shows the attack surface extending beyond traditional cybersecurity boundaries. A software update may address a network-facing flaw while changing behavior that matters under sensor noise, low-glucose conditions, connectivity loss, or patient-specific patterns. A test suite may demonstrate that the vulnerability no longer triggers, but not that clinical safety properties remain within acceptable bounds. Here accountable partial understanding requires that security evidence not be isolated from safety, clinical, and operational evidence, because an attacker or failure mode that satisfies one evidence domain while violating another exploits precisely the gap between them.

Across all three examples, the adversarial significance of partial understanding is the same. The attacker benefits when the organization cannot distinguish claim from evidence, authority from intent, explanation from justification, or uncertainty from acceptance. This is why accountable partial understanding is a security requirement rather than an epistemic preference: it treats the limits of understanding as something attackers exploit and defenders must design around, part of the system rather than a property of an outside knower.

\section{What must be understood, and by whom}
\label{sec:forms}

The argument so far leans on the word \emph{understanding}, which can easily become too broad. Software security uses the term in many ways. A reverse engineer understands a binary by recovering structure and behavior. A developer understands a vulnerability by tracing a flawed assumption through code. An incident responder understands an intrusion by reconstructing an adversary's path. A verifier understands a system by proving that a specified property holds within a model. An operator understands a defensive action by knowing what happened, why it happened, and what to do next. These are related, but they are not the same, and distinguishing them sharpens what agentic systems change.

\subsection{Traditional forms}

Traditional software-security work has emphasized three related forms of understanding.

\emph{Structural understanding} concerns what a system is made of and how its parts relate. It asks what components exist, which dependencies are present, what interfaces connect them, where trust boundaries lie, and which assets, identities, privileges, and data flows matter. In practice it appears in architecture diagrams, dependency graphs, attack-surface maps, software bills of materials, data-flow diagrams, and threat models. It gives defenders a map of the system they are trying to protect.

\emph{Behavioral understanding} concerns what the system does under specified conditions. It asks how software responds to inputs, state, configuration, failures, and adversarially chosen interactions. Testing, fuzzing, symbolic execution, runtime monitoring, simulation, and formal verification all produce behavioral evidence, though with different coverage and strength. Behavioral understanding is what lets a defender say, within some scope, that a program rejects invalid input, enforces an access-control rule, preserves an invariant, or fails safely.

\emph{Causal understanding} concerns why a particular behavior, vulnerability, or failure occurred. It asks what chain of events, assumptions, interactions, or design choices produced the outcome. It appears in root-cause analysis, incident reconstruction, vulnerability analysis, forensic timelines, and postmortems. Defenders need to understand how bad behavior became possible, not merely that it occurred.

These forms remain central, and nothing in this paper argues for abandoning them. Accountable partial understanding depends on them, because a system cannot be accountable if no one can identify its components, observe its behavior, or reconstruct important causal chains. But structural, behavioral, and causal understanding do not by themselves answer the questions that agentic systems raise. They tell defenders what the system is, what it did, and why an outcome occurred. They do not necessarily tell anyone what authority was exercised, what evidence justified action, who accepted the residual uncertainty, or how a human could meaningfully intervene.

\subsection{Forms that delegated autonomy demands}

Three further forms become load-bearing when software acts under delegated authority. None is new in itself. Operational awareness has mattered since the first operator watched a console; assurance has an institutional history running through the Common Criteria and national evaluation partnerships~\cite{commoncriteria}; forensics and incident response grew precisely to serve accountability. What changes in the agentic era is who produces these forms of understanding, at what rate, and through what mediation, because the four shifts of Section~\ref{sec:agentic} make each of them harder to obtain by traditional means at the moment they are needed most.

\emph{Operational understanding} concerns what the system is doing now, what state it has placed the organization in, and what a human should do next. This is the understanding needed by a SOC analyst responding to an alert, an operator supervising an autonomous defensive action, or a clinician encountering a changed software workflow. It is time-sensitive, role-specific, and oriented toward action. A complete architecture diagram or after-the-fact explanation cannot provide it. During an anomaly or degraded mode, the relevant question is whether the human can form an adequate mental model quickly enough to intervene, override, defer, escalate, or accept the system's action~\cite{endsley1995situation}.

\emph{Assurance understanding} concerns what evidence supports bounded trust in the system. It asks not only whether the system behaved as claimed but what kind of evidence supports the claim, what scope the evidence covers, what assumptions it depends on, and what remains outside it. Test results, proofs, audit logs, simulations, assurance cases, red-team findings, and provenance records all contribute. The key distinction is between evidence and assertion. A model-generated summary or confidence score may be useful input, but assurance understanding requires that claims connect to evidence whose scope and limitations are themselves inspectable.

\emph{Accountability understanding} concerns who can justify, challenge, approve, reverse, or accept responsibility for a security-relevant action. It asks who authorized an autonomous agent to act, who can contest the action, who maintains the evidence behind it, who owns the assumptions on which it depended, and who accepts the residual unknowns. This form matters most when authorship and authority are distributed across humans, models, tools, pipelines, and organizations. A system may be structurally documented, behaviorally tested, and causally reconstructed, yet still fail accountability if no actor can say who was responsible for a decision, what evidence justified it, or how it could have been challenged before harm occurred.

These forms shift the security question from whether the system can be analyzed to whether the partial results of analysis reach the right actors, at the right time, in a form that supports action. For the coding agent, structural understanding of the codebase and behavioral evidence from tests matter, but the organization also needs to know what assumptions the patch relied on, who accepted the change, and what uncertainty remains. For the cyber-defense agent, causal reconstruction after an incident matters, but the operator needs operational understanding \emph{while} the agent is acting. For the insulin delivery update, behavioral tests matter, but assurance understanding must connect cybersecurity claims to safety, clinical function, and regulatory expectations simultaneously.

The boundary between these families is porous. A useful audit trail may contribute to causal, assurance, and accountability understanding at once; a well-designed operator interface may combine structural cues, behavioral predictions, and recommended actions. The point is that software security cannot rely solely on the forms of understanding that made sense when software was primarily an artifact to be inspected. When software also acts, explains, and authorizes, understanding must extend to evidence, authority, responsibility, intervention, and residual uncertainty.

\section{Accountable partial understanding}
\label{sec:apu}

The phrase \emph{accountable partial understanding} (APU) is deliberately composed of three parts. It is \emph{partial} because full comprehension is unavailable or unrealistic. It is \emph{understanding} because security still requires intelligible claims about structure, behavior, evidence, authority, and action. It is \emph{accountable} because the limits of those claims must be made visible, assigned, and actionable rather than hidden behind automation, fluency, or organizational ambiguity.

\subsection{Definition}

Accountable partial understanding is a security property of systems that cannot be fully comprehended by any single human or organization, in which the limits of understanding are explicitly bounded, supported by inspectable evidence, tied to assumptions, assigned to responsible actors, maintained over time, and made actionable for intervention, audit, or contestation.

Different stakeholders need different understandings for different decisions, and some uncertainty will remain. A developer may need to know what code changed and why. An operator may need to know whether an autonomous response is still within delegated authority. An auditor may need to know what evidence supports a claim. A patient, customer, or affected party may need to know what decision affected them and how it can be challenged. Accountable partial understanding is therefore a set of relationships among claims, evidence, assumptions, actors, authority, and action, not a single artifact, interface, or explanation.

Partial understanding names a condition, not a goal. The normative claim is that when partial understanding is unavoidable, security practice should make it explicit and governable. Security value comes from knowing which part is understood, which evidence supports that understanding, what remains unknown, and what forms of intervention remain possible.

\subsection{Merely partial versus accountable partial}

Partial understanding by itself is ordinary. Every deployed system is understood only partially by any individual user, developer, operator, vendor, or regulator. A developer may understand one subsystem but not the deployment environment. A security analyst may understand an alert but not the model that ranked it. An operator may understand that an agent changed a firewall rule but not why that change was selected over alternatives. A regulator may understand the submitted evidence but not the full operational context in which the system will later run.

Merely partial understanding becomes dangerous when it is mistaken for sufficient understanding. This happens when approval workflows obscure who accepted residual uncertainty, when a model-generated explanation substitutes for evidence, when passing tests are treated as evidence of properties the tests do not cover, or when logs record that an action occurred but not the authority or assumptions behind it. In such cases, the organization possesses fragments of understanding without knowing how those fragments relate to decisions, risks, or responsibilities.

Accountable partial understanding adds structure to those fragments. It requires that partial claims be accompanied by answers to a specific kind of question: what exactly is claimed, what evidence supports the claim, what assumptions must hold, who is responsible for accepting or challenging it, what remains unknown, what actions are available, and what changes would invalidate the understanding. The contrast can be stated simply. Merely partial understanding says that some things about the system are known. Accountable partial understanding says \emph{which} things are known, on what evidence, under what assumptions, for whose decision, with what residual unknowns, and with what mechanisms for action.

One clarification guards against overreading. APU converts some unknown unknowns into known unknowns that can be named, bounded, monitored, and acted upon. It does not, and cannot, eliminate unknown unknowns. Adversaries retain surprises; zero-day vulnerabilities remain possible; the register of unknowns is itself partial. If APU eliminated the unknown, there would be no security problem left to manage. The claim is narrower and more useful: an organization that names and bounds what it does not know presents fewer unguarded gaps between claim and evidence than one that does not, and it can respond faster when a surprise arrives. Nor does APU eliminate or transfer risk. It disciplines how residual uncertainty is represented and owned so that risk decisions rest on something firmer than assertion.

\subsection{Necessary conditions}

For partial understanding to be accountable, six conditions must be satisfied. These are not a compliance checklist. They are design commitments, properties that make incomplete understanding visible, governable, and useful for security decisions.

\paragraph{Bounded scope and explicit assumptions.}
The scope of understanding must be explicit, and the assumptions on which it depends must be stated. A claim may apply to a component, model, dependency, configuration, workflow, policy, threat model, or operating environment, but it should not silently exceed its scope. A test suite provides evidence about selected behaviors, not all behaviors. A proof establishes a property within a formal model, not outside the model's assumptions. A vulnerability patch may assume a particular input-sanitization path; a defensive action may assume that an observed login is malicious; a medical-device update may assume a sensor behavior or control-loop condition. If the boundaries and assumptions are hidden, the limits of the understanding are hidden as well. Bounded scope prevents a local understanding from being mistaken for a global one; explicit assumptions make it possible to recognize when the understanding has been invalidated.

\paragraph{Inspectable evidence.}
Claims about system behavior should connect to evidence that can be inspected independently of the agent or tool that produced the claim. Evidence may include source diffs, test results, traces, logs, proofs, simulations, provenance records, red-team findings, review comments, or structured assurance arguments. The important distinction is between assertion and evidence. An AI system may help produce evidence, organize evidence, or point humans toward evidence, but its assertion that a system is safe, patched, compliant, or low risk is not itself sufficient evidence.

\paragraph{Assigned responsibility.}
Partial understanding becomes accountable only when responsibility for claims and decisions is assigned. A single person need not understand or own the whole system. Instead, some person or organizational role must be responsible for accepting the evidence, maintaining the assumption set, approving the delegated authority, responding to challenges, or accepting residual uncertainty. Without assigned responsibility, distributed authorship becomes distributed unaccountability.

\paragraph{Visible residual unknowns.}
Accountable partial understanding requires explicit representation of what remains unknown, including untested behaviors, unverified properties, unavailable logs, unexplained model behavior, opaque dependencies, and operational conditions that lie outside the evidence. Residual unknowns are not embarrassing gaps to be hidden. They are security-relevant facts that decision makers need in order to accept, defer, escalate, or constrain action. The observation that concealing unknowns was never good practice is correct, and APU does not claim to have discovered honesty. What it adds is mechanism and obligation: unknowns must be recorded in a form that is connected to specific claims, owned by specific actors, and capable of triggering specific responses, rather than living in the professional judgment of whoever happened to notice them.

\paragraph{Actionability.}
Understanding must support action by the relevant actor. A developer needs information that supports review or repair. An operator needs information that supports intervention, override, escalation, or rollback. An auditor needs information that supports evaluation. An affected party may need information that supports contestation. A collection of evidence that no relevant actor can interpret or act upon may be documentation, but it is not accountable understanding.

\paragraph{Temporal maintenance.}
Understanding must be maintained as systems change. A claim that was well supported yesterday may be invalidated by a dependency update, model change, prompt change, configuration change, policy change, environmental shift, or newly discovered adversarial technique. This matters most for agentic systems because the same workflow that produces an action may also modify the artifacts on which prior understanding depended. Accountable partial understanding therefore requires mechanisms for recognizing when previous evidence, assumptions, or approvals no longer apply.

Together, these conditions turn partial understanding from a passive state into an active practice. They do not guarantee security, they do not remove uncertainty, and other properties such as resilience, fail-safe behavior, and graceful degradation may also be needed in specific domains. They make uncertainty visible and connected to evidence, responsibility, and action.

\subsection{Accountability to whom?}
\label{sec:towhom}

A common failure of transparency and explainability efforts is the assumption that a single explanation can serve everyone. In security, the relevant audiences differ sharply. As Table~\ref{tab:audiences} shows, developers, operators, incident responders, auditors, affected users, and executives may all need understanding, but they do not need the same understanding, at the same time, or for the same purpose.

\begin{table}[tbp]
  \centering
  \caption{Different actors require different forms of accountable partial understanding.}
  \label{tab:audiences}
  \begin{tabular}{p{0.24\linewidth}p{0.66\linewidth}}
    \toprule
    \textbf{Actor} & \textbf{Needed understanding} \\
    \midrule
    Developer &
    What changed, why it changed, what assumptions the change relies on, and what evidence supports the change. \\
    \addlinespace
    Operator &
    What the system is doing now, whether it remains within delegated authority, and how to intervene, escalate, or roll back. \\
    \addlinespace
    Incident responder &
    What happened, what evidence supports the reconstruction, what state the system is now in, and what actions remain available. \\
    \addlinespace
    Auditor or regulator &
    What claims are being made, what evidence supports them, what scope and assumptions limit them, and who accepted residual uncertainty. \\
    \addlinespace
    Affected user, patient, or customer &
    What decision or action affected them, what recourse or contestation path exists, and what information can be disclosed safely. \\
    \addlinespace
    Executive or procurer &
    What residual unknowns are being accepted, what accountability mechanisms exist, and what risks cannot be transferred to vendors or tools. \\
    \bottomrule
  \end{tabular}
\end{table}

This role-specific view is central to accountable partial understanding. It avoids two common mistakes. The first is assuming that more information automatically produces better understanding. Too much information, delivered to the wrong actor at the wrong time, increases cognitive burden while leaving security unimproved. The second is assuming that a high-level explanation is enough. A natural-language summary may help an executive or operator form an initial orientation, but a developer may need a diff, a test, and a trace; an auditor may need a structured argument with explicit scope and assumptions; and an incident responder may need provenance, timestamps, and system state.

A third consideration deserves emphasis. None of these actors arrives empty. Each brings prior knowledge, professional training, past incidents, and a self-image, and these shape what any accountability view can accomplish. Understanding is not delivered onto a blank slate; it is integrated into a mental model the actor already holds. An experienced analyst may discount evidence that contradicts a favored hypothesis. A practitioner whose professional identity rests on expertise may resist artifacts that foreground what remains unknown, because security culture has often rewarded the appearance of certainty over the honest answer ``it depends.'' Role-specific views should therefore be designed with the recipient's likely priors, expertise, and tolerance for uncertainty in mind, and organizations adopting APU should expect the cultural work of making uncertainty speakable to be as demanding as the technical work of recording it. Accountability in security does not mean being all-knowing; it means owning a defined role in a system no one fully knows.

Every actor need not receive every detail. In many security contexts, full disclosure may be impossible or unsafe. The point is that each actor should receive the form of understanding needed to fulfill their role, partial but accountable to evidence and action.

\subsection{What APU is not}

APU can be misunderstood in several ways, and distinguishing it from adjacent concepts clarifies its scope. Section~\ref{sec:paradigms} develops these relationships in depth; four brief contrasts orient the reader now.

Risk acceptance operates at a different decision point. Risk management asks what risks should be accepted, mitigated, transferred, or avoided under uncertainty. APU asks how the uncertainty itself is represented, evidenced, assigned, monitored, and made actionable. Risk acceptance without APU can become a ritual in which organizations sign off on risk registers without knowing precisely what uncertainty they are accepting or what would change their assessment.

Explainable AI contributes at most one class of artifact. Explanations may support APU, but explanation is one artifact among many, and the relevant object is broader than a model output; it is a security-relevant system of models, tools, authority, evidence, assumptions, and human roles. A fluent explanation disconnected from inspectable evidence may reduce understanding.

Formal verification and assurance cases provide disciplined evidence, but they do not remove partiality. Formal methods provide strong evidence for specified properties within defined models and assumptions; assurance cases organize claims, evidence, and argumentation. Both discipline partial understanding by making the scope of what is known explicit and rigorous. Neither eliminates partiality, and both must be maintained as systems change. APU treats these methods as instruments, not as escapes from the condition they help govern.

Human-in-the-loop designs satisfy APU only when the human has evidence, timing, and authority sufficient to intervene. A human who receives an opaque recommendation, a generated summary, or an approval button without sufficient evidence or authority cannot supply accountability by presence alone~\cite{bainbridge1983ironies}.

\section{Design principles for accountable partial understanding}
\label{sec:principles}

If unaccountable partial understanding is an attack surface, then accountable partial understanding must be designed. Post-hoc explanations cannot carry the burden, and organizations cannot assume a human reviewer will infer what matters from a generated summary. The system must be structured so that autonomous actions remain bounded, evidenced, reviewable, and contestable even when full understanding is unavailable.

The five principles below translate the conditions of Section~\ref{sec:apu} into design commitments. They are not a complete methodology; their purpose is to make the limits of understanding concrete enough to be engineered, tested, and governed. Readers familiar with secure development lifecycle practice will recognize ancestors of each principle: least privilege, independent review, configuration management, provenance, and change control appear throughout frameworks such as the NIST Secure Software Development Framework~\cite{nist2022ssdf}. The principles here do not replace those controls. They extend them to conditions the frameworks largely assume away: artifacts authored by agents, change arriving at machine pace, and an evidence base that the acting system can itself modify.

\subsection{Bounded delegation}

Autonomous and AI-mediated systems should operate within explicit bounds on what they may decide, modify, invoke, deploy, or authorize. Delegation should be treated as a security property, not a workflow convenience. A system that summarizes code requires different authority than a system that edits code. A system that recommends a firewall change requires different authority than one that deploys it. A system that drafts a safety-critical update requires different authority than one that releases it to patients.

Bounded delegation requires more than access control in the narrow sense. It requires an explicit account of which actions the system may take, under what conditions, with what evidence, and with what review or rollback requirements. Permission to use a tool matters only in relation to the type and quality of understanding available at the time, and permission should degrade gracefully, so that the scope of delegated authority narrows as evidence becomes thinner or assumptions become less certain. Bounded delegation also requires that the delegating entity, whether human or agent, can inspect what it is delegating. When agents delegate to other agents, the chain of authority must roll up to a human in a form that remains interpretable. A record that delegation occurred is not enough; the record must represent what was authorized, under what conditions, and what accountability was preserved at each step in the chain.

Setting these bounds raises the question of who decides them. The answer cannot be the agent at run time. Which actions are reversible, which require pre-action review, and which decisions are non-delegable are themselves security decisions that accountable humans must make in advance, record as delegation policy, and revisit as evidence and conditions change. The agent operates within the policy; it does not author it.

For the coding agent, bounded delegation may allow patch generation and test proposals while requiring independent review before merging security-sensitive changes. For the cyber-defense agent, it may allow rapid temporary containment while requiring operator confirmation before persistent policy changes. For the insulin delivery system, it may mean separating automated vulnerability discovery from deployment decisions that affect clinical behavior.

\subsection{Evidence-producing and decoupled actions}

Autonomous actions with security consequences should produce evidence as part of acting. A system should not only produce a patch, recommendation, classification, or policy change. It should also preserve the artifacts needed to evaluate the action, including the inputs considered, the assumptions made, the tools invoked, the alternatives rejected, the tests or analyses performed, the uncertainty encountered, and the authority exercised.

This principle distinguishes evidence from explanation. A natural-language explanation may help a human understand an action, but it cannot carry the evidentiary burden alone. The explanation should point to artifacts that can be inspected independently, such as diffs, traces, logs, test results, proof obligations, provenance records, review comments, and assurance arguments. AI can help produce and organize such evidence, but the claim that an action was correct, safe, or low risk should not rest solely on the AI's assertion.

Evidence production alone fails, however, when the same agent that acts also controls the evidence for that action. If an agent produces a patch, generates the tests, summarizes the results, and certifies their sufficiency, accountability collapses into a polished form of self-attestation. Evidence-producing actions should therefore be \emph{decoupled}; the mechanism that evaluates a claim should be independent, to the extent possible, of the workflow that produced it.

Decoupled attestation can take several forms. \emph{Artifact diversity} means checking an AI-generated patch against a pre-existing regression suite, deterministic static analyzer, formal property, or independently maintained test harness that the agent did not author or modify. \emph{Process separation} means ensuring that the agent proposing a change does not also control the approval policy, logging configuration, test oracle, or residual-unknown register on which accountability depends. \emph{Adversarial review} means using an independently governed tool, model, human review process, or red-team procedure to attempt to invalidate the claim, identify missing assumptions, or expose gaps in scope.

The independence need not be perfect to be useful, but it must be meaningful. Two models using the same prompts, tools, data sources, and organizational incentives provide redundancy without genuine decoupling. The more consequential the delegated action, the stronger the required separation should be.

\subsection{Assumption tracking, invalidation, and dynamic probing}

Accountable partial understanding requires that assumptions be treated as first-class security artifacts. Security claims depend on assumptions about environment, configuration, dependencies, user behavior, threat model, data quality, model behavior, and operating conditions. When these assumptions are implicit, partial understanding appears stronger than it is. When they are explicit, they can be reviewed, challenged, monitored, and invalidated.

Assumption tracking matters most for agentic systems because their behavior may depend on context that is transient or difficult to reconstruct, such as retrieved documentation, prior interaction history, tool outputs, prompts, or environmental state. Assumptions should therefore be attached to the claims and evidence they support. A test result should indicate what behavior it exercises and what it does not. A summary should indicate what sources it considered and what it may have omitted. A delegated action should record the conditions under which its authority was taken to apply. The aim is visibility of critical assumptions, not exhaustive capture of every background belief.

Tracking assumptions is necessary but not sufficient. APU also requires mechanisms for recognizing when prior understanding no longer applies. Software systems change continuously as dependencies update, models are replaced, prompts are edited, tools are added, configurations drift, policies change, operating environments shift, and adversaries adapt. Each change can invalidate prior evidence, assumptions, approvals, or delegation boundaries. Agentic systems raise the stakes because they may modify the artifacts that support their own accountability. A coding agent may update tests with code, a defense agent may alter logging or policy, and a summarization workflow may change what evidence humans see. Without evidence preservation and invalidation rules, yesterday's well-founded understanding becomes today's unwarranted confidence.

These properties should not merely be asserted; they should be \emph{probed}. An organization should be able to test whether assumption tracking and invalidation work under change and adversarial pressure. One probe dimension is \emph{evidence linkage}, which asks whether each claim behind a given action can be traced to inspectable evidence, explicit assumptions, and a responsible actor. A second is \emph{invalidation latency}, which asks how quickly the system recognizes that prior understanding no longer applies. Auditors can introduce invalidation probes, such as dependency updates, configuration drift, changed tool permissions, missing telemetry, and altered prompts, and observe whether the system surfaces affected claims, identifies which evidence has been invalidated, narrows delegated authority if necessary, and updates the residual unknown register. Dynamic probing makes accountability a temporal security property, not a procedural ritual.

\subsection{Role-specific accountability views}

As Table~\ref{tab:audiences} illustrates, different actors need different forms of understanding. The design implication is that accountable partial understanding should not be a single artifact or interface but a set of views, each tailored to the action the recipient is expected to take and to the knowledge the recipient already holds.

This principle addresses a common failure of transparency, in which too much information is delivered without regard to role, timing, or actionability. More information differs from more understanding. A SOC analyst in the middle of an incident does not need an exhaustive record of every intermediate reasoning step; they need enough operational understanding to decide whether to accept, override, escalate, or reverse an action. An auditor, by contrast, may need a more complete evidentiary record after the fact. A clinician affected by a device software change may need a different view than the manufacturer's security team. Accountable partial understanding requires all of these, but not in the same form, at the same time, or at the same level of detail.

\subsection{Intervention-preserving design and residual unknown registers}

Autonomous systems should be designed so that meaningful human intervention remains possible when it matters. Intervention is meaningful only if the human has the authority, time, evidence, and interface needed to act. A human who can only click approve or reject on an opaque recommendation lacks meaningful oversight. A human who receives evidence only after an irreversible action has occurred cannot prevent harm.

Intervention-preserving design requires attention to timing and reversibility. Some actions can be automated safely because they are low impact, temporary, or easily reversed. Others should require pre-action review, staged deployment, limited duration, or explicit human authorization. Classifying actions into these categories, and deciding when intervention ``matters,'' is part of the delegation policy that accountable humans set in advance under bounded delegation; it is not a judgment the agent makes for itself. The system should make clear when it is acting autonomously, when it is requesting approval, when it is escalating uncertainty, and when it has reached the boundary of its delegated authority. The test is whether the loop gives the human a genuine position rather than a ceremonial one.

Meaningful intervention requires knowing where understanding ends, and \emph{residual unknown registers} make that boundary explicit. A residual unknown register records what remains unverified, unexplained, untested, unmodeled, unavailable, or outside the scope of current evidence. It may state that a patch was tested against known exploit paths but not against all related input paths, that a defensive action was based on incomplete telemetry, or that a safety-critical update has security evidence but awaits additional clinical validation.

Residual unknowns are not defects in accountable partial understanding; they are among its most important outputs. Registers should connect to action, so that unknowns trigger additional testing, limit deployment, require escalation, constrain delegated authority, or inform risk acceptance. They may also expire. An unknown tolerable during emergency containment may be unacceptable for long-term operation. The aim is a living account of uncertainty that prevents unknowns from being silently converted into assumed understanding.

\medskip

These five principles make partial understanding visible, evidenced, bounded, and actionable without pretending to restore full comprehension. They are interdependent. Bounded delegation is meaningful only if the evidence behind it is decoupled from the agent that acts; evidence is useful only if the assumptions it depends on are tracked; assumptions are governable only if invalidation is recognized and probed; and intervention is meaningful only if residual unknowns are visible. Accountable partial understanding lies in the discipline of maintaining all of these together.

\section{Relationship to existing approaches}
\label{sec:paradigms}

Accountable partial understanding draws from several established traditions, including risk management, assurance cases, formal methods, explainable AI, and research on trust in automation and human factors. Each addresses part of the problem. None is sufficient by itself for AI-mediated and autonomous software systems that act under delegated authority.

\subsection{Risk management}

Risk management begins from the premise that uncertainty cannot be eliminated. Organizations identify risks, estimate likelihood and impact, select mitigations, and decide which residual risks to accept, avoid, transfer, or monitor. In this sense, APU aligns with risk management.

The difference is that risk management treats uncertainty as an input to decision-making, while APU asks how that uncertainty is represented, evidenced, assigned, and made actionable inside the system and workflow~\cite{singer2020trust}. A risk register may record that a vulnerability, dependency, model behavior, or operational condition presents residual risk. APU asks whether the organization knows what evidence supports that entry, what assumptions bound it, who owns it, what would invalidate it, and what actions are available if the uncertainty becomes operationally important.

This distinction matters for autonomous systems because risk acceptance can become detached from the evidence needed to make acceptance meaningful. An executive may accept a risk described in summary form. An operator may later confront the same risk as an ambiguous system state. A developer may be asked to patch it with partial knowledge of its causes. An auditor may need to reconstruct whether the accepted risk matched the system as deployed. These actors are connected in principle but rarely in practice; the interconnections that would make a risk register a shared instrument are exactly what APU supplies. It connects risk acceptance to the evidence, assumptions, roles, and intervention mechanisms that make acceptance more than a procedural act.

\subsection{Assurance cases and safety cases}

Assurance cases and safety cases provide the closest existing precedent for APU. They organize claims, arguments, and evidence to justify confidence that a system satisfies particular properties in a particular context~\cite{kelly2004gsn, mohamad2021securityassurance}. They make explicit that assurance is more than a collection of tests or documents; it is a structured argument about why evidence supports a claim. APU borrows this evidence-centered orientation directly: it asks what claim is being made, what evidence supports it, what assumptions limit it, and what remains outside its scope.

The difference lies in the setting in which the argument must be maintained. Traditional assurance cases assume a relatively stable artifact, a relatively stable body of evidence, and a human-managed change process. AI-mediated autonomous systems challenge all three assumptions. The artifact may change through generated code, updated prompts, tool changes, model updates, or adaptive workflows. The evidence may be selected, summarized, or produced by AI-mediated tools. The change process may operate at a pace that makes manual assurance maintenance difficult.

Assurance cases are therefore not obsolete under APU; they are among its most important mechanisms. But they must become more dynamic, role-specific, and adversarially aware. Prior work has examined the relationship between trust and assurance in socio-technical systems, arguing that the two must be carefully distinguished~\cite{flechais2005divide}. APU extends this distinction to settings where the system itself helps produce the artifacts under assurance. Assurance arguments for agentic systems should track delegated authority, generated evidence, assumption changes, residual unknowns, and invalidation events, the concerns developed in Sections~\ref{sec:apu} and~\ref{sec:principles}. APU can be understood as a broader framework within which assurance cases serve as one instrument for disciplining partial understanding over time.

\subsection{Formal methods}

Formal methods provide some of the strongest available tools for making claims about software behavior precise. Model checking, theorem proving, type systems, and runtime verification reduce ambiguity and provide evidence stronger than testing alone for the properties they cover~\cite{baier2008principles}.

For APU, the central contribution of formal methods lies in making the scope of understanding explicit. A proof depends on a property, a model, assumptions, abstractions, and a correspondence between the formal artifact and the deployed system. Those boundaries are not weaknesses; they are the conditions under which formal evidence is meaningful. Verification does not eliminate partial understanding; it disciplines it. This is precisely what APU values, namely evidence whose scope and limitations are themselves inspectable.

The relationship grows more complex when AI mediates the formal process. AI may assist in generating specifications, candidate invariants, tests, or proof obligations. Such assistance can lower the barrier to rigorous evidence and extend its reach to systems that would otherwise go unverified. But AI assistance does not make a system verified. A model may suggest an invariant that a verifier can check, or summarize a proof obligation that a human can inspect for relevance. The central distinction remains. AI can help \emph{produce} formal evidence, but the evidence must be \emph{independently} inspectable. Generated assertion differs from verified claim.

\subsection{Explainable AI}

Explainable AI (XAI) addresses a related but narrower problem: making model outputs, classifications, recommendations, or behaviors interpretable to humans~\cite{lipton2018mythos, ribeiro2016whytrust}. This matters for AI-mediated security systems, especially when models triage alerts, summarize incidents, rank vulnerabilities, or recommend actions.

APU subsumes explainability while asking a broader question. XAI typically asks whether a model's output can be made interpretable; APU asks whether a \emph{system's} security-relevant behavior, spanning models, tools, authority, evidence, assumptions, and human roles, can be made accountable. Put another way, APU demands not explainable AI but justifiable security action, in which an explanation counts only when it connects to inspectable evidence, is scoped to a role, and is presented with its limitations. A fluent explanation that substitutes for evidence, or that serves one actor's needs while leaving others without actionable understanding, does not satisfy APU even if it satisfies XAI.

\subsection{Trust in automation and human factors}

Research on trust in automation has long recognized that effective automation depends on calibrated human reliance~\cite{lee2004trust}. Humans over-trust automation, under-trust it, lose situation awareness, or are asked to intervene after their understanding of the system has degraded. These concerns bear directly on autonomous software security.

APU builds on this tradition by adding an explicitly adversarial and evidentiary frame. Calibrated trust is necessary, but security requires more than a well-calibrated mental model. The distinction between trustworthiness and trust matters here: a system may be genuinely trustworthy yet fail to earn human trust, or humans may extend trust to a system that is not trustworthy. APU addresses both directions. By making the grounds for trustworthiness explicit and inspectable, it provides a basis for calibrating human trust rather than leaving that calibration to impressions formed under adversarial pressure. Attackers manipulate the information on which trust is based, exploiting alert fatigue, misleading summaries, ambiguous authority, and gaps in provenance. APU therefore asks whether, beyond calibrated trust, the system provides role-appropriate evidence, preserves intervention, records assumptions, and makes residual unknowns visible even under adversarial pressure.

This also clarifies why ``human in the loop'' is insufficient as a security principle. Human involvement improves accountability only when the human has the authority, evidence, time, and interface needed to act meaningfully. Otherwise the human becomes a \emph{liability sink}, formally present in the workflow and absorbing organizational responsibility while not positioned to understand, challenge, or govern the system's behavior. The cyber-defense agent that presents an operator with an ``approve'' button and a generated summary, under time pressure, with no visibility into the evidence or alternatives, is an example of liability sinking, not meaningful oversight. APU reframes human oversight as a design requirement, in which the human must be equipped for the understanding their role demands, rather than a procedural checkbox.

\medskip

Across these traditions the pattern is consistent. Risk management addresses uncertainty, assurance cases organize evidence, formal methods make claims precise, explainable AI supports intelligibility, and human-factors research addresses reliance and intervention. APU does not replace these traditions. It connects them around the specific problem of securing autonomous and AI-mediated software systems when full understanding is unavailable, partial understanding is unavoidable, and adversaries can exploit the gaps between them.

\section{Open questions}
\label{sec:open}

Accountable partial understanding is a framework, not a completed method. The principles in Section~\ref{sec:principles} describe what security practice should try to preserve when full understanding is unavailable, but each raises difficult research questions. They cluster around three concerns: evidence and measurement, responsibility and authority, and resilience under adversarial pressure.

\subsection{Evidence and measurement}

How can accountable partial understanding be measured? Security practice already measures many proxies, such as test coverage, vulnerability counts, mean time to remediate, alert volume, audit findings, model accuracy, and compliance status. These do not directly measure whether partial understanding is accountable. A system might have high test coverage but weak assumption tracking, extensive logs but poor reconstructability, or accurate model outputs but no clear path for human contestation. Candidate measures include assumption coverage (what fraction of critical assumptions are explicitly recorded), evidence linkage (whether claims are connected to inspectable artifacts), action reconstructability (whether an action can be traced to its inputs, authority, and evidence), rollback readiness, residual unknown visibility, and stakeholder actionability. Each is difficult to operationalize. An empirical research program might begin by examining existing AI-mediated security workflows, such as coding agents, SOC copilots, and automated remediation pipelines, and assessing which of these properties are present, absent, or unmeasured. Without some way to assess accountable partial understanding, it risks remaining an appealing concept instead of an operational security property.

A closely related question concerns the role of AI in producing security evidence. The circularity problem is unavoidable because AI systems help generate tests, propose invariants, summarize incidents, draft assurance arguments, identify assumptions, and organize provenance. The important distinction is between AI as a producer of independently inspectable artifacts and AI as the authority that certifies those artifacts. An AI system's assertion that its own output is correct, safe, or sufficiently evidenced cannot be treated as sufficient assurance. Research is needed on where this boundary should be drawn, how independent checking should work, what forms of AI-generated evidence are trustworthy under what conditions, and how to prevent AI-generated evidence from becoming a more polished form of self-attestation.

\subsection{Responsibility, authority, and orchestration}

How should accountability be allocated across humans, models, tools, vendors, deployers, and organizations? Agentic workflows distribute authorship and authority. An action with security consequences may involve a foundation model provider, an application developer, a vendor-supplied tool, an organizational policy, a human operator, and an automated deployment pipeline. Existing accountability mechanisms assume clearer boundaries than these workflows provide. APU requires more precise ways to assign responsibility for claims, assumptions, evidence maintenance, delegated authority, and residual risk acceptance. It also requires ways to prevent responsibility from diffusing across the socio-technical system until no actor can meaningfully answer for the result, the distributed unaccountability problem identified in Section~\ref{sec:apu}.

Partiality also implies parts, and parts require orchestration. If different actors hold different partial understandings connected by evidence, someone or something must compose those parts into a security-relevant whole by deciding which views exist, how they connect, and who maintains the composition itself. Orchestration is its own locus of responsibility and its own attack surface. Whether it can be partially automated without recreating the self-attestation problem, and who answers for the orchestrator, are open questions.

These questions connect to a deeper one about the boundary between security and governance. In autonomous systems, questions of authority, evidence, contestability, and responsibility are not merely administrative. They shape what the system is allowed to do, how its actions can be challenged, and whether humans can intervene before or after harm occurs. Governance becomes part of the security mechanism. That raises design questions with no purely technical answers. Which decisions must remain non-delegable? Which residual unknowns can be accepted by an operator, and which require executive, regulatory, or affected-party involvement? When should a system refuse to act because it cannot produce sufficient evidence for accountable action? Nor is governance politically neutral. Accountability regimes are shaped by commercial competition, national rivalry over AI capability, and liability pressure, and each creates incentives to under-record unknowns or over-claim evidence. These are security questions when software acts with delegated authority, and they require collaboration among security engineers, system designers, legal scholars, policy researchers, and domain experts.

\subsection{Adversarial resilience}

Section~\ref{sec:attack-surface} argued that unaccountable partial understanding is an attack surface. The open question is how accountability mechanisms should themselves be defended. Attackers may target both software behavior and the artifacts that make behavior accountable, by poisoning retrieved context, manipulating logs, inducing misleading summaries, generating evidence that satisfies shallow checks, or overloading human reviewers with too many plausible changes. Designing for APU therefore requires threat models that include attacks on evidence, assumptions, explanations, provenance, delegation boundaries, and human attention.

A related challenge is maintaining accountable partial understanding as systems change. A model update, prompt change, dependency update, new tool permission, generated patch, or modified workflow may invalidate prior evidence. Section~\ref{sec:principles} addressed assumption tracking and invalidation as a design principle, but the mechanisms for detecting invalidation, identifying which claims are affected, and triggering appropriate review remain largely unspecified. The hard problem lies less in producing assurance evidence once than in maintaining the relationship among claims, evidence, assumptions, and authority over time, particularly when the system under assurance is also an agent that may modify its own evidence base. Dynamic assurance for mutable, AI-mediated systems is perhaps the most technically demanding of these open problems.

These questions form the agenda that follows from taking partial understanding seriously as a design condition. If full understanding is unavailable, the central question for security research becomes how to make incomplete understanding bounded, evidenced, assigned, actionable, and resilient against the adversaries who will test its limits.

\section{Conclusion}
\label{sec:conclusion}

Software security has always operated under incomplete understanding, and attackers have always worked in the gap. What AI-mediated and autonomous systems change is the character of the gap. Partial understanding becomes faster-moving, more distributed, more opaque, and more tightly coupled to delegated authority. Software is no longer only the object that defenders attempt to understand. It is also a participant in producing the code, evidence, explanations, and actions that defenders are asked to trust. Under those conditions, the limits of understanding stop being background and become an attack surface that adversaries can probe and expand.

Accountable partial understanding reframes the defensive problem accordingly. Instead of treating uncertainty as an embarrassing residue to be hidden, it treats uncertainty as a condition that demands visibility, assignment, and governance. Instead of treating the limits of understanding as an outside knower's epistemic problem, it treats those limits as part of the system, something to be bounded, evidenced, owned, probed, and defended. APU does not demand that every actor understand the whole system. It demands that partial understandings be explicit enough, evidenced enough, and actionable enough to support security decisions and withstand adversarial pressure.

The design principles derived here do not restore a lost world of complete comprehension; no such world existed. Their intent is more modest and more practical. When understanding is partial, its scope should be bounded, its evidence inspectable, its assumptions tracked, its authority assigned, and its gaps known; the principles make each of those properties an explicit object of design. They also open research questions about measurement, dynamic assurance, AI-assisted evidence, responsibility allocation, orchestration, and the boundary between security and governance. And although the running examples are software-centric, nothing confines APU to software development. Security operations, procurement, incident response, and policy all make consequential decisions under the same conditions, and the same discipline applies. Name what is known, evidence it, own it, and keep the unknowns in view.

Security without full understanding is already here. The task is to make it accountable.

\section*{Acknowledgements}

The author thanks the anonymous reviewers of an earlier version of this paper for comments that substantially sharpened its framing and argument.

\bibliographystyle{unsrtnat}
\bibliography{bibliography}

\end{document}